\documentstyle[12pt,epsf,aaspp4]{article}
\def\gsim{\;\rlap{\lower 2.5pt
 \hbox{$\sim$}}\raise 1.5pt\hbox{$>$}\;}
\def\lsim{\;\rlap{\lower 2.5pt
   \hbox{$\sim$}}\raise 1.5pt\hbox{$<$}\;}

\newcommand{\lya}{Ly$\alpha$ }

\newcommand{\beq}{\begin{equation}}
\newcommand{\eeq}{\end{equation}}
\def\myputfigure#1#2#3#4#5%
{\hskip0.03\textwidth\vskip#5pt
\makebox[0pt]{\hskip#2in
\includegraphics[width=#3\textwidth]{#1}}\vskip#4pt\hfill}


\lefthead{Haiman}
\righthead{High Redshift Ly$\alpha$ Emission Lines Prior to Reionization}

\begin{document}
\title{The Detectability of High Redshift Ly$\alpha$ Emission Lines \\
Prior to the Reionization of the Universe}
\author{Zolt\'an Haiman\altaffilmark{1}} \affil{Princeton University
Observatory, Princeton, NJ 08544, USA\\ zoltan@astro.princeton.edu}
\vspace{0.5\baselineskip}
\altaffiltext{1}{Hubble Fellow}

\vspace{\baselineskip}
\begin{abstract}
For a source of Ly$\alpha$ radiation embedded in a neutral
intergalactic medium (IGM) prior to the reionization epoch, the
emission line is strongly suppressed by the intervening IGM.  The
damping wing of the so--called ``Gunn-Peterson trough'' can extend to
the red side of the emission line, and erase a significant fraction of
the total line flux.  However, the transmitted fraction increases with
the size of the local cosmological HII region surrounding the source,
and therefore with the ionizing luminosity and age of the source.
Motivated by the recent discovery of a Ly$\alpha$ emitting galaxy at a
redshift $z=6.56$ (Hu et al. 2002), possibly prior to the reionization
of the IGM, we revisit the effects of a neutral IGM on the Ly$\alpha$
emission line.  We show that even for faint sources with little
ionizing continuum, the emission line can remain observable.  In
particular, the line detected by Hu et al. is consistent with a source
embedded in a neutral IGM.  We provide characterizations of the
asymmetry and total transmitted flux of the Ly$\alpha$ line as
functions of the ionizing emissivity of its source.  A statistical
sample of Ly$\alpha$ emitters extending beyond the reionization
redshift can be a useful probe of reionization.

\end{abstract}

\newpage
\section{Introduction}
\label{sec:introduction}

Observations of high redshift quasars out to $z\sim 6$, and the lack
of a strong HI absorption trough (the so--called ``Gunn--Peterson '',
GP, trough) have revealed that the intergalactic medium (IGM) is
highly ionized between redshifts $0\lsim z \lsim 6$.  On the other
hand, observations of the temperature anisotropies of the cosmic
microwave background (CMB) radiation, and the lack of a strong damping
of these anisotropies by electron scattering, have shown that the
universe was neutral between the redshifts $30\lsim z \lsim 10^3$.  It
is natural to identify the first generation of galaxies or quasars as
the sources of ultraviolet (UV) radiation that reionized the
intergalactic medium at some redshift $z_r$ between $6\lsim z_r \lsim
30$ (e.g. Haiman \& Loeb 1998). How and when reionization happened,
and the details of the reionization process, have been among the most
pressing outstanding questions in astrophysical cosmology, likely
holding many clues about the first generation of light sources.

The bright quasar SDSS 1030+0524 recently discovered in the Sloan
Digital Sky Survey (SDSS) at redshift $z=6.28$ has revealed the first
GP trough, i.e., a spectrum consistent with no flux at a substantial
stretch of wavelength shortward of $(1+z)\lambda_\alpha=8850\AA$.  The
lack of any detectable flux implies a strong lower limit $x_{\rm
H}\gsim 0.01$ on the mean mass--weighted neutral fraction of the IGM
at $z\sim 6$ (Fan et al. 2002).  Furthermore, the neutral fractions
inferred from a sample of high redshift quasars from $5\lsim z \lsim
6$ show a rapid rise towards high redshift. Comparisons with numerical
simulations of cosmological reionization (Cen \& McDonald 2002; Gnedin
2002; Fan et al. 2002) show that the IGM is likely neutral at
$z\gsim6.3$.

It has long been suggested that the first sources of ultraviolet
radiation, responsible for the reionization of the universe, could be
strong Ly$\alpha$ emitters (see the review by Pritchet 1994 and
references therein).  Indeed, numerous Ly$\alpha$ sources have
recently been discovered at high redshifts (Dey et al. 1998; Weymann
et al. 1998; Spinrad et al. 1998; Hu et al. 1999; Ellis et al. 2001;
Rhoads \& Malhotra 2001; Malhotra \& Rhoads 2002). These discoveries
demonstrate that Ly$\alpha$ emitting galaxies do exist in significant
numbers out to the current horizon of their detectability at $z\sim
6$, despite the potentially strong susceptibility of Ly$\alpha$
photons to dust absorption (Charlot \& Fall 1993). The observed space
densities are broadly consistent with ``semi--analytical'' models that
incorporate redshift-dependent dust obscuration, and connect
Ly$\alpha$ emitting galaxies and dark matter halos (Haiman \& Spaans
1999).  

The present paper is motivated by the recent discovery of a Ly$\alpha$
emitting galaxy at a redshift of $z=6.56$ (Hu et al. 2002), which is
possibly {\it prior} to the reionization redshift of the IGM.  It has
been pointed out that Ly$\alpha$ photons injected into a neutral IGM
are strongly scattered, and that the red damping wing of the GP trough
can strongly suppress, or even completely eliminate the Ly$\alpha$
emission line (Miralda-Escud\'e 1998; Miralda-Escud\'e \& Rees 1998;
see also Loeb \& Rybicki 1999 for a more detailed treatment of the
scattering process).  The reionization of the IGM may therefore be
accompanied by a rapid decline in the observed space density of
Ly$\alpha$ emitters beyond $z_r$ (Haiman \& Spaans 1999). Indeed, such
a decline could provide a useful observational probe of the
reionization epoch in a large enough sample of Ly$\alpha$ emitters
(Haiman \& Spaans 1999; Rhoads \& Malhotra 2001), complementary to
methods utilizing the GP trough.

Naively, the two recent observations mentioned above present a puzzle:
a Ly$\alpha$ emission line is observed prior to the reionization
epoch, where it should have been strongly suppressed.  The simplest
resolution of this puzzle is obtained by considering the cosmological
HII region created by the Ly$\alpha$ source itself.  As shown by Cen
\& Haiman (2001) and Madau \& Rees (2001), a source with a bright
ionizing continuum can create a large ($\gsim 30$ comoving Mpc) cosmological
HII region.  For a sufficiently luminous source, the size of the HII
region corresponds to a wavelength range $\Delta\lambda$ that exceeds
the width of the emission line, allowing most of the intrinsic
Ly$\alpha$ line to be transmitted without significant scattering.

Motivated by the apparent recent observational puzzle, we here revisit
this problem, and consider the effect of the IGM on the intrinsic
Ly$\alpha$ emission line of a {\it faint} galaxy, such as the one
discovered by Hu et al.  (2002).  The ionizing continuum for this
source is less than $\sim 10^{-3}$ of that of the bright $z\sim 6$
SDSS quasars (Fan et al. 2002). We show that even for faint sources
with such little ionizing continuum, a significant fraction of the
emission line can remain observable.  In particular, we argue that the
Ly$\alpha$ line detected by Hu et al. (2002) is consistent with a
source embedded in a neutral IGM.

The rest of this paper is organized as follows.  In \S~2, we briefly summarize
our model used to predict the line profile. In \S~3, we apply this
model to the $z=6.56$ Ly$\alpha$ emitter.  In \S~4, we study how the
shape of the transmitted Ly$\alpha$ line depends on our model
parameters.  In \S~5, we discuss our results and offer our
conclusions. Throughout this paper, we assume the background cosmology
to be flat $\Lambda$CDM with $(\Omega_\Lambda,\Omega_{\rm
m},\Omega_{\rm b}, h)=(0.7,0.3,0.04,0.7)$.

\section{Transmission of the \lya Emission Line}
\label{sec:transmission}

In this section, we describe our model for the effects of the IGM on
the observed \lya emission line profile.  This discussion closely
parallels the treatment in Cen \& Haiman (2001), and is only briefly
summarized here.

Consider an ionizing source at redshift $z_\star$ with a given
(steady) star formation rate $\dot M_\star$ and age $t_\star$. We
assume the source is embedded in an initially neutral IGM with a mean
density of $\langle\rho_{\rm IGM}\rangle=\Omega_{\rm b}\rho_{\rm
crit}(1+z_\star)^3$, and a mean gas clumping factor of $C\equiv\langle
\rho_{\rm IGM}^2 \rangle / \langle \rho_{\rm IGM} \rangle^2$.  
We first compute the size of the ionized region around the source.
Ignoring any recombinations, the radius of the ``Str\"omgren sphere''
is $R_{\rm s}= 0.6 (\frac{M_\star}{40~{\rm M_\odot~yr^{-1}}})^{1/3}
(\frac{t_\star}{10^7~{\rm yr}})^{1/3} (\frac{1+z_\star}{7.56})^{-1}$ 
(proper) Mpc,
where we have used the fact that over its lifetime, a stellar
population with a Salpeter initial mass function (IMF) extending from
0.1 to 120 ${\rm M_\odot}$ produces $\approx 4000$ ionizing photons
per stellar proton, and equated the number of photons to the
number of hydrogen atoms inside $R_s$. This is accurate for low
clumping factors and source ages, but recombinations can decrease the
size $R_s$ for ($C\ga 10$, or $t_\star\ga 10^8$ yrs). The results
presented below are obtained by a numerical solution that accounts for
recombinations.

For simplicity, we assume further that the source emits a Ly$\alpha$
line with a thermally broadened profile, so that the unobscured flux
at observed wavelength $\lambda$ is $F_{\rm \lambda,0} \propto
\exp\{-[\lambda-(1+z_\star)\lambda_\alpha]^2/\Delta\lambda^2\}$,
where $\lambda_\alpha=1215\AA$, and $\Delta\lambda=
(1+z_\star)\lambda_\alpha\Delta v/c$ is the apparent Doppler
width at thermal velocity $\Delta v$.  The observed line profile 
at wavelength $\lambda$ will then be
$F_{\rm \lambda,obs} = F_{\rm \lambda,0}
\exp(-\tau_\lambda)$, where $\tau_\lambda=-\int_0^{z_\star} dz (cdt/dz)
n_{\rm H} \sigma_{\lambda/(1+z)}$ is the optical depth to Ly$\alpha$
absorption.  In the last integral, $n_{\rm H}$ is the neutral hydrogen
number density along the line of sight to the source, and
$\sigma_{\lambda}$ is the Ly$\alpha$ cross--section (assumed here to
be given by a Voigt profile, using temperatures of $10^4$K and
$2.728(1+151)[(1+z)/151)]^2$K inside and outside the HII region,
respectively).

The intrinsic Ly$\alpha$ emission line is reprocessed by the opacity
of the intervening neutral IGM, as well as by the residual neutral
hydrogen within the Str\"omgren sphere.  We here assume that outside
the HII region, the IGM is fully neutral, and $n_{\rm H}(z)=
0.76\rho_{\rm IGM}(z)/m_p$.  Inside the HII region, we explicitly compute
the neutral fraction as a function of redshift (or distance from the
source), by equating the ionizing flux to the recombination rate,
assuming that the latter is boosted relative to a uniform IGM by the
value of the gas clumping factor $C$.  The value of the clumping
factor, and the shape of the density distribution, have little effect
on the transmission on the red side of the line, but it does effect
the transmission on the blue side of the line (see discussion below).

\section{The \lya Line of the Galaxy at $z=6.56$}
\label{sec:huetal}

The galaxy HCDM 6A was found in a narrow--band imaging survey behind
the cluster Abell 370 with the LRIS camera on the Keck 10m telescopes
(Hu et al. 2002).  It is an intrinsically faint source, and was
detected through its gravitational lensing magnification by a factor
of $\sim 5$.  Broad--band optical and near--infrared imaging was used
to infer a star formation rate of 40 ${\rm
M_\odot~yr^{-1}}$. Subsequent LRIS spectroscopy revealed a Ly$\alpha$
line, with the inferred redshift of $z_\star=6.56$.  The star
formation rate inferred from the line using Kennicutt's (1983) law is
2 ${\rm M_\odot~yr^{-1}}$, approximately $\sim20$ times smaller than
implied by the continuum flux.  The fact that the emission line is
underluminous may indeed be an indication that a fraction of the line
was scattered by the neutral IGM (although there are other plausible
explanations, such as dust absorption, or a large escape fraction of
ionizing photons from the galaxy, as discussed below).

Taking the observed parameters of this source, we now use the model in
the previous section to predict the transmitted shape of the emission
line.  We adopt the star--formation rate of $\dot M_\star=40~{\rm
M_\odot~yr^{-1}}$ as inferred from the observed continuum (translating
to an ionizing photon production rate of $\dot N_{\rm ph}=6\times
10^{54}~{\rm s^{-1}}$).  We adopt an intrinsic thermal line--width of
$\Delta v = 230~{\rm km~s^{-1}}$ and source redshift of
$z_\star=6.556$; as we will see below, this pair of choices roughly
reproduces the width (FWHM $\sim 15$\AA), and peak wavelength
($\lambda_{\rm max}\sim 9187$\AA) of the observed line once the IGM
opacity is taken into account.  We further assume an age of
$t_\star=10^7$ yr for the source, a gas clumping factor of $C=10$, and
an escape fraction for the ionizing photons of $f_{\rm esc}=10\%$.
The resulting size of the HII region under these assumptions is 0.3
proper Mpc (or 2 comoving Mpc). Reasonable variations on the assumed
parameters do not change our main conclusions, as discussed below.

\begin{figure}
\plotone{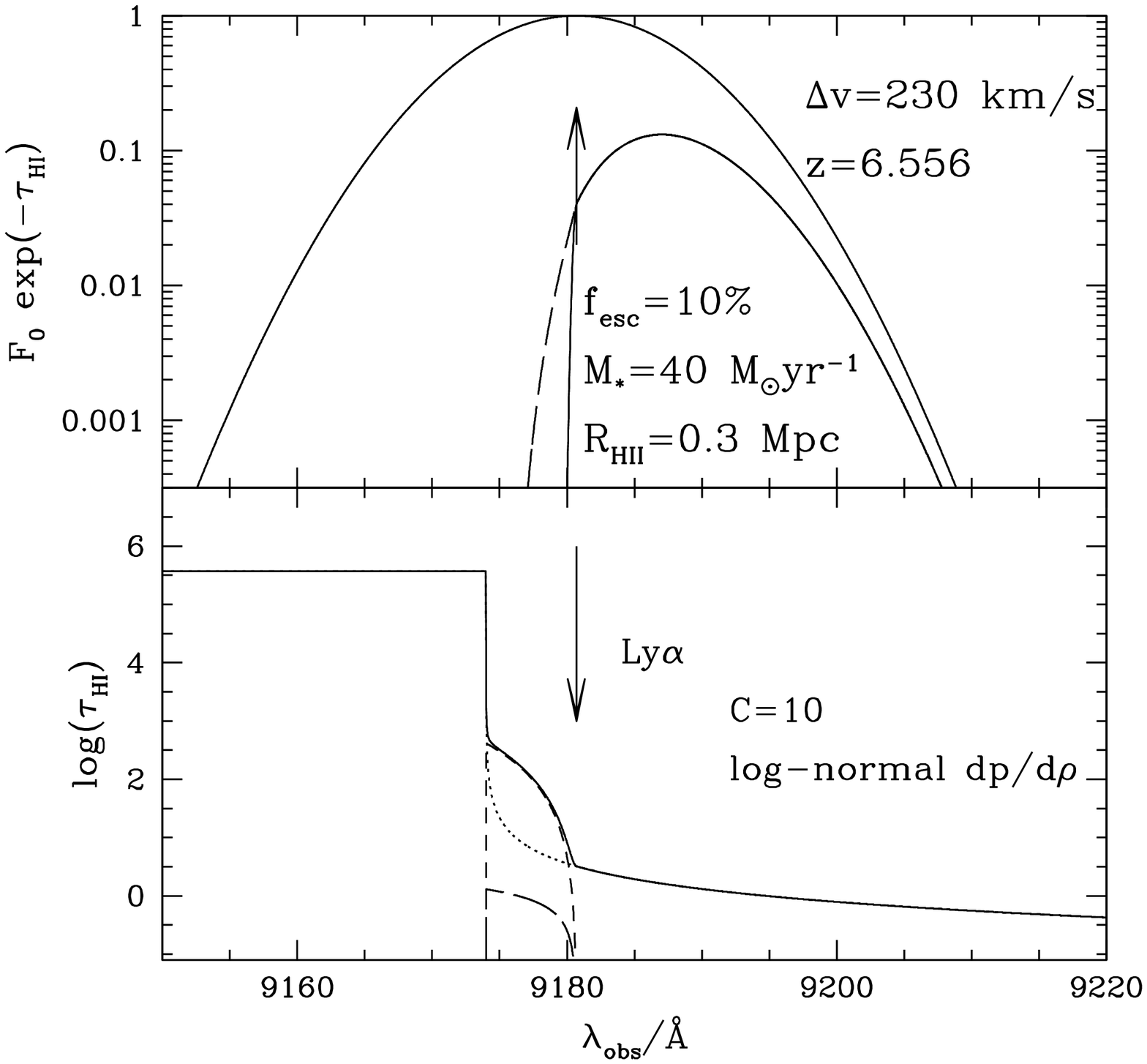}
\caption{{\it Upper Panel:} Profile of the \lya line from a $z=6.56$
galaxy. The top curve shows the adopted intrinsic profile, and the bottom curve
shows the profile including absorption in the IGM and by the neutral atoms
inside the $0.3$ Mpc HII region surrounding the source.  {\it Lower Panel:} The
optical depth as a function of wavelength from within the HII region (short and
long dashed curves, corresponding to different treatments of the HI opacity
within the HII region; see text for discussion), from the neutral IGM outside
the HII region (dotted curve), as well as from the sum of the two (solid
curve).  The arrows in both panels indicate the central wavelength of the
unobscured Ly$\alpha$ line.}
\label{fig:spec}
\end{figure}

The upper panel of Figure~\ref{fig:spec} shows the resulting profile
of the \lya emission line. The top solid curve shows the assumed
unabsorbed line profile, normalized to have a peak flux of unity (in
arbitrary units).  The bottom solid curve shows the profile including
the opacities of both the IGM and the neutral atoms within the HII
region.  In the lower panel of Figure~\ref{fig:spec}, we show the
separate contributions to the total HI opacity (solid curve) from
within the HII region (short-dashed curve) and from the neutral IGM
(dotted curve).

Figure~\ref{fig:spec} reveals that at the center of the Ly$\alpha$
line, the opacity is only of order unity ($\tau_{\rm tot}\approx 3$).
On the blue side of the line, the opacity is dominated by the residual
neutral hydrogen inside the HII region, and is significantly higher,
while on the red side of the line, the opacity is dominated by the
damping wing of the IGM, and falls below unity.  The most important
point revealed by this figure is that the opacities are not
exceedingly large, and a significant fraction of the intrinsic line
flux is transmitted.  The peak flux is reduced by a factor of $\sim
10$ relative to its unabsorbed value, and $\sim 8\%$ of the total
intrinsic line flux is transmitted.  This is consistent with the
finding by Hu et al. that the line is underluminous by a factor of
$\sim 20$ compared to what is expected from the continuum, and allows
an additional suppression of the total line flux by $\sim 60\%$
(e.g. by dust).

We have so far implicitly assumed that the transmitted flux equals
$F=F_0\exp(-\langle \tau_{\rm HI}\rangle)$, where $F_0$ is the
unobscured flux, and $\langle \tau_{\rm HI}\rangle$ is the mean HI
opacity.  This should be a reasonable assumption for the damping wing
of the IGM, since a large range of redshifts $\Delta z\sim 0.1-0.2$
contribute to the opacity at any given wavelength. Density
fluctuations along the line of sight will therefore average out to
result in the mean opacity.  However, on the blue side of the line,
the opacity at a given observed wavelength is dominated by a narrow
redshift interval, corresponding to the thermal width ($\sim 10$ km/s)
around the redshift at which the photons pass the through the
Ly$\alpha$ resonance.  As a result, depending on the spectral
resolution, there could be significant opacity fluctuations within a
single spectral resolution element.  In this case, the observed flux
would be better approximated by $F=F_0\langle \exp(-\tau_{\rm
HI})\rangle$, where the average is taken over the probability
distribution of $\tau_{\rm HI}$.  This follows directly from the
density distribution, which we here assume is log--normal (which has
only one free parameter, $C$), together with the fact that within the
HII region, where ionization equilibrium is rapidly established,
$\tau_{\rm HI}\propto \rho^2$.

In general, $\langle \exp(-\tau_{\rm HI})\rangle$ can be much larger
than $\exp(-\langle \tau_{\rm HI}\rangle)$, implying that we may have
significantly {\it underestimated} the transmission on the blue side
of the line. To illustrate this effect, in Figure~\ref{fig:spec} we
show the line profile using $\tau_{\rm eff}=-\ln \langle
\exp(-\tau_{\rm HI})\rangle$ as the effective opacity of the neutral
hydrogen inside the HII region.  The long-dashed curve in the lower
panel shows that this reduces the opacity by a factor of $\sim 100$,
and that the damping wing of the IGM now dominates the opacity across
the entire profile.  The long--dashed curve in the upper panel show
the transmitted line shape.  The reduced opacity makes the line appear
more symmetric, and slightly increases the total transmitted flux.

We also note that ``extreme'' gas clumping could reduce the
transmitted flux, by reducing the size of the HII region around the
source.  This could be the case if the galaxy HCDM 6A is situated in a
large--scale density condensation.  In particular, one may assume that
the galaxy is part of a larger halo with an NFW (Navarro, Frenk \&
White 1997) density profile, in which the gas has an extended
distribution with a hot and cold component in pressure
equilibrium. Under these assumptions, it would require implausibly
large star formation rates to ionize the gas in the halo, and there
would be no cosmological HII region outside the local halo (see Haiman
\& Rees 2001).  The fact that the galaxy HCDM 6A does show Ly$\alpha$
emission suggests that it is not surrounded by such a large scale gas
distribution, i.e. most of the gas had to cool and assemble into a
disk, or into neutral clumps with a low covering factor.

\section{Properties of Partially Obscured Ly$\alpha$ Lines}
\label{sec:properties}

To address how the results of the previous section depend on the
parameters of our model, Figure~\ref{fig:sfr} shows various properties
of the processed Ly$\alpha$ emission line for sources with different
star formation rates. Note that a change in the lifetime of the source
(until its Str\"omgren sphere approaches its equilibrium size), or in
the escape fraction of ionizing photons, is essentially equivalent to
changing the star formation rate, since what matters most for the
transmitted spectrum is the overall size of the HII region.  We use an
emission line at $z=6.56$ with an unobscured line--width of $230~{\rm
km~s^{-1}}$, and a pure Gaussian intrinsic profile, as an illustrative
example; and we also set $f_{\rm esc}=1$.

As the SFR is decreased, the source becomes fainter, and larger parts
of the Ly$\alpha$ line are obscured.  This is demonstrated on the four
panels of Figure~\ref{fig:sfr}.  The upper left panel shows that as
the SFR is decreased, the central wavelength of the line appears to
shift towards the red, by upto $\approx 10 \AA$.  The upper right
panel shows that the flux at the apparent peak of the line decreases
by upto a factor of $\sim 20$.  Interestingly, even in the absence of
any HII regions (e.g. for a very faint source), the transmitted flux
at the apparent line center is still $\sim 5\%$ of the peak flux of
the unobscured line.  In the lower left panel, we show the
``asymmetry'' in the line profile, which we define as the ratio of the
transmitted fluxes on the blue vs. red side of the apparent
line-center.  Apparently, as the size of the HII region is decreased,
the line becomes first increasingly asymmetric. However there is a
maximum in the asymmetry at a star formation rate of around $100~{\rm
M_\odot~yr^{-1}}$, and for still lower emissivities, the line would
appear increasingly symmetric again.  Finally, the lower right panel
shows how the total transmitted line flux depends on the star
formation rate.  The total line flux decreases by upto a factor of
$\sim 30$, but interestingly, even in the absence of any HII region,
the total transmitted line flux still contains $\sim 3\%$ of its
unobscured value.

The dashed lines in Figure~\ref{fig:sfr} correspond to the case in
which gas clumping with a log--normal density distribution inside the
HII region reduces the effective opacity, as discussed in the previous
section.  This does not effect the location or height of the observed
peak, and only slightly increases the total transmitted flux. However,
it does significantly reduce the asymmetry of the line.

The assumed intrinsic width of the Ly$\alpha$ line can significantly
effect the above results. In Figure~\ref{fig:width}, we show how the
apparent line center, peak flux, asymmetry, and total line flux depend
on the assumed width $\Delta v$.  The set of curves marked by vertical
lines correspond to a model with $f_{\rm esc}=100\%$ and a
corresponding proper HII region size of 0.6 Mpc.  The other set of
curves describe the parameters of the Ly$\alpha$ line in a model
without any HII region.  The Figure shows that intrinsically broader
lines are less effected by the IGM opacity.  In particular, the lower
solid curve in the bottom right panel indicates that for a line
broader than $\gsim 1000$ km/s, $\sim 10\%$ of the intrinsic flux is
transmitted even in the absence of any HII region.  Conversely, for
intrinsic line widths significantly narrower than $\lsim 200$ km/s,
significant flux can remain only if the source is surrounded by an HII
region whose size is at least $\sim 0.1$ (proper) Mpc.

\section{Discussion and Conclusions}
\label{sec:conclude}

We have shown that a significant fraction of the intrinsic Ly$\alpha$
line of a galaxy should be detectable, even for faint galaxies
embedded in a neutral IGM.  This implies that contrary to previous
suggestions, Ly$\alpha$ emitting galaxies should be observable from
beyond the reionization redshift.  In particular, the emission line
detected by Hu et al. is consistent with a galaxy embedded in a
neutral IGM.  Using the star formation rate of 40 ${\rm
M_\odot~yr^{-1}}$ inferred from the continuum of this source, and
assuming an escape fraction of ionizing photons $f_{\rm esc}=10\%$,
and age of $10^7$ years, we find that it should be surrounded by an
ionized region of size $\sim 2$ (comoving) Mpc. In this case, $\sim
10\%$ of the intrinsic line flux is transmitted, consistent with the
measured line strength if dust, or local neutral gas, has absorbed 
an additional $\sim 50\%$ of the line.

\begin{figure}
\plotone{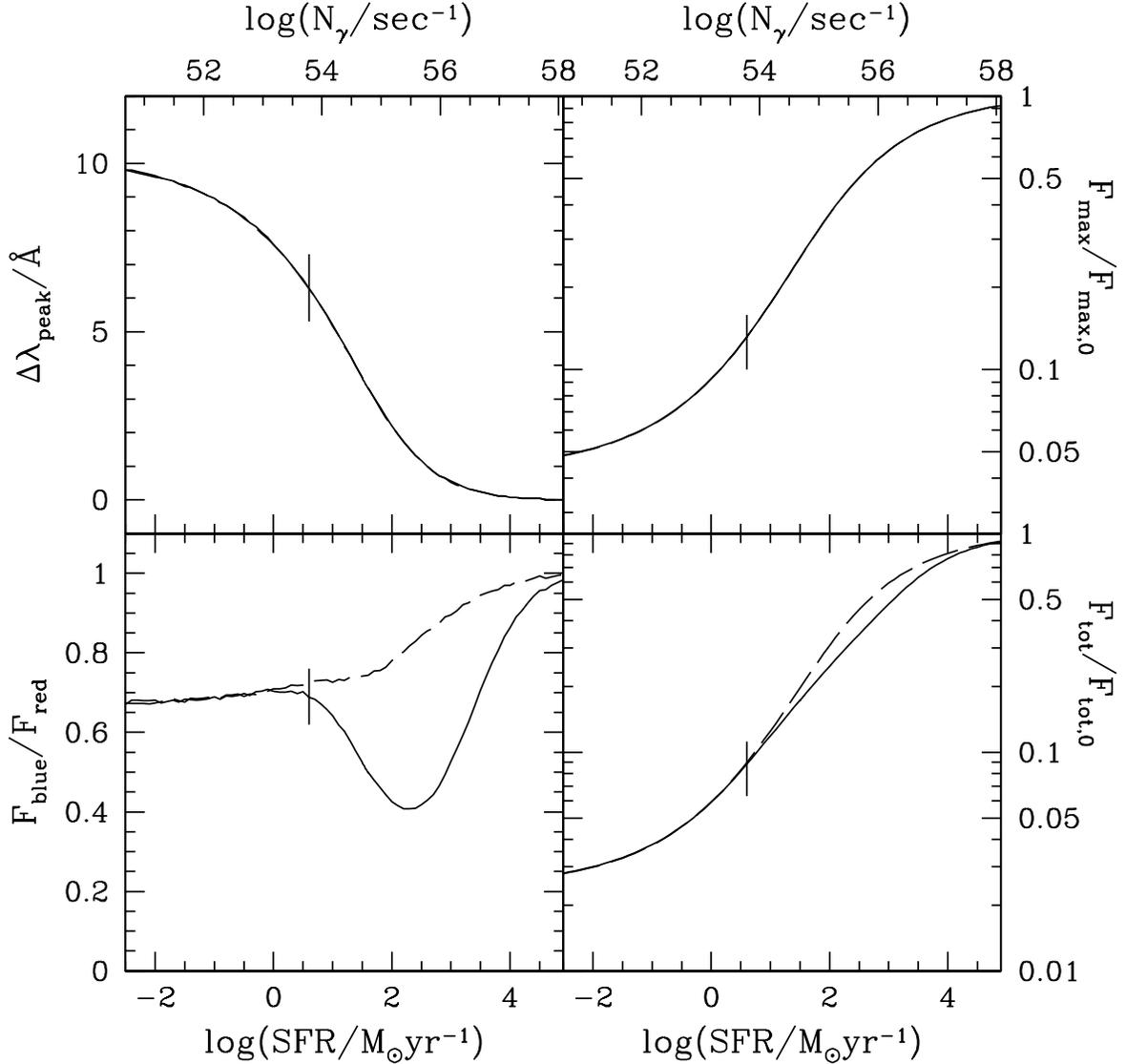}
\caption{The parameters of the observed \lya line as a function of 
the star formation rate of the galaxy at $z=6.56$.  The intrinsic line
is assumed to have a width of $230~{\rm km~s^{-1}}$.  The four panels
show, clockwise from top left corner: (a) the apparent redward
displacement of the line center; (b) the suppression of the apparent
peak flux relative to the unobscured line; (c) the apparent asymmetry
of the line (ratio of fluxes on the apparent blue and red sides); and
(d) the suppression of the total line flux relative to the unobscured
line.  The solid and dashed curves correspond to different treatments
of the HI opacity within the HII region (see text for discussion).}
\label{fig:sfr}
\end{figure}

\begin{figure}
\plotone{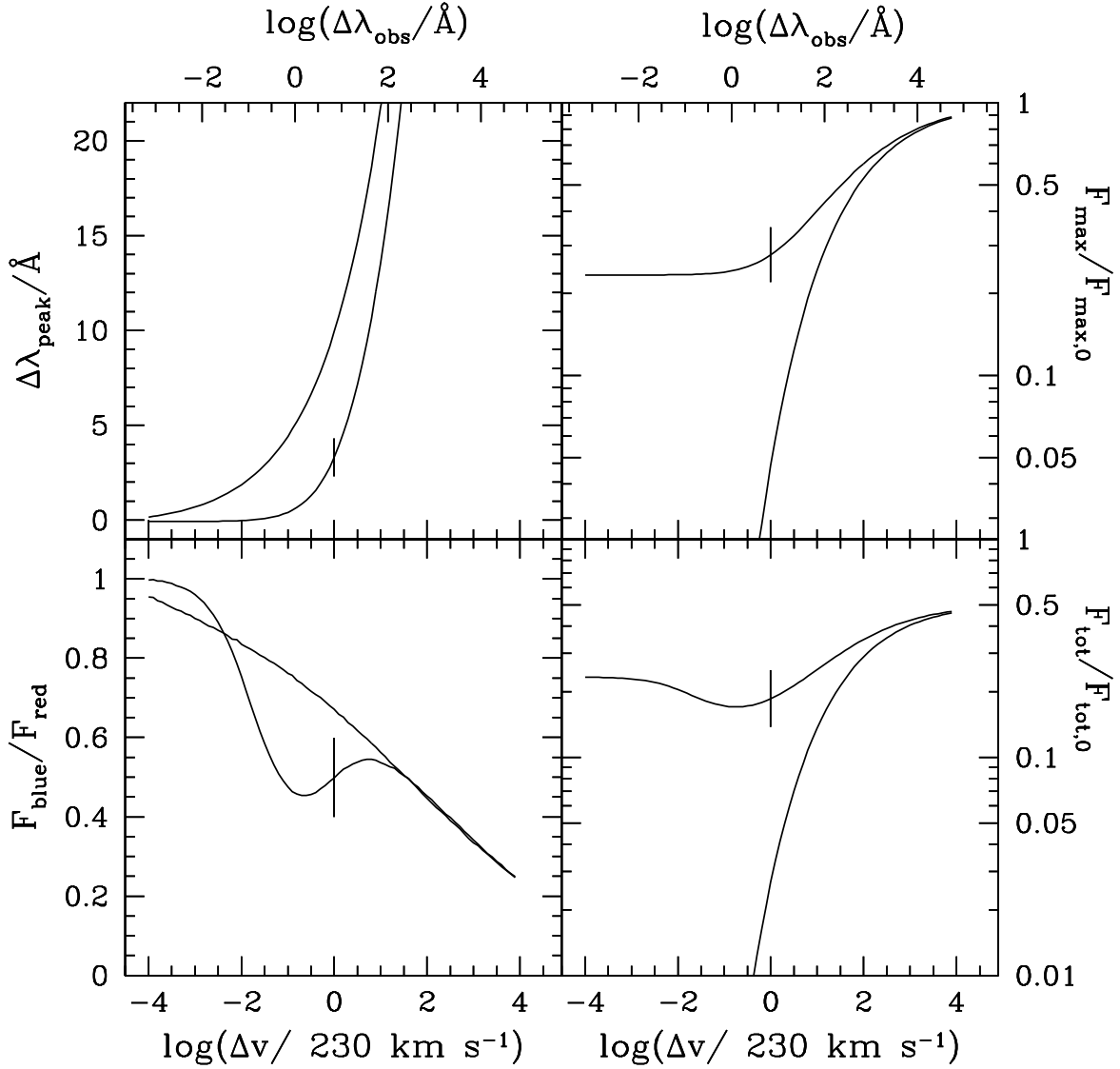}
\caption{The parameters of the observed \lya line as a function of the 
assumed intrinsic line width. The four panels show the same quantities
as Figure~\ref{fig:sfr}. The set of curves marked by vertical lines
correspond to a model with $f_{\rm esc}=100\%$ and a proper HII region
size of 0.6 Mpc.  The other set of curves describe the parameters of
the Ly$\alpha$ line in a model without any HII region.}
\label{fig:width}
\end{figure}

In general, the observed line profiles for sources prior to
reionization should have properties that correlate with the size of
the local HII region, and therefore with the luminosity of the source.
For an illustrative example of a thermal emission line with a width of
$230~{\rm km~s^{-1}}$ at redshift $z=6.56$, we have demonstrated how
the observed shape depends on the star formation rate of the source.
The results are shown in Figure~\ref{fig:sfr}, and also strongly
depend on the intrinsic line--profile, as shown in
Figure~\ref{fig:width}. For example, for a line as narrow as $\sim 20$
km/s, the maximum apparent redward displacement of the line--center
(in the absence of an HII region) would be $\sim 5\AA$, rather than
$\sim 10\AA$, and for a star formation rate of $\lsim 1$ ${\rm
M_\odot~yr^{-1}}$, the line would essentially be erased, with less
than a fraction of $10^{-3}$ of the total line flux transmitted. On
the other hand, for a line as broad as $\sim 2000$ km/s, the maximum
displacement could be upto $\sim 20\AA$, and $\gsim 30\%$ of the total
flux would be transmitted even for arbitrarily faint sources.  The
parameters of the transmitted line shape for other values of the
assumed intrinsic linewidth and SFR can be read off Figures
\ref{fig:sfr} and \ref{fig:width}.

We conclude that relatively faint Ly$\alpha$ emitting galaxies should
be observable from epochs prior to the reionization of the universe.
A statistical sample of Ly$\alpha$ emitters that spans the
reionization redshift should be a useful probe of reionization,
through the study of the correlations between the luminosity of the
sources and the properties of the emission lines, such as their total
line/continuum ratio (if a continuum is measured), the asymmetry of
the line profile, and the offset of the peak of the line from the
central Ly$\alpha$ wavelength (for sources that have redshift
measurements from other emission lines).  These studies will serve as
a useful complement to probing the reionization of the IGM through
observations of the GP trough in the spectra of distance quasars, or
through observations of the CMB anisotropies.

\acknowledgements

The author thanks James Rhoads and Renyue Cen for useful comments and
discussions.  This work was supported by NASA through the Hubble
Fellowship grant HF-01119.01-99A, awarded by the Space Telescope
Science Institute, which is operated by the Association of
Universities for Research in Astronomy, Inc., for NASA under contract
NAS 5-26555.

\begin{figure}
\plotone{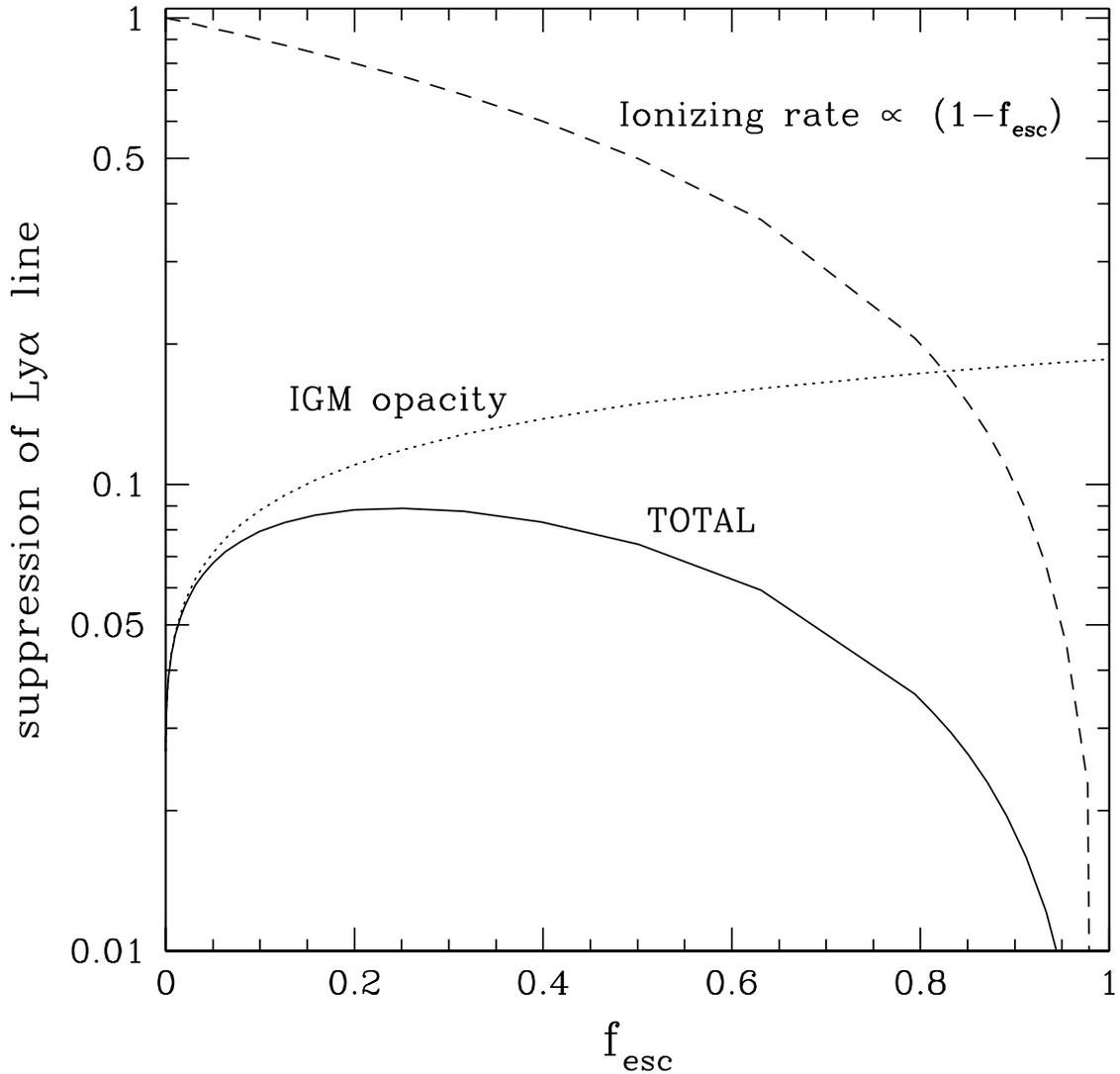}
\caption{The overall suppression of the Ly$\alpha$ line flux as a 
function of the escape fraction of ionizing radiation (solid curve).
This is a product of two factors: only a fraction $(1-f_{\rm esc})$ of
the ionizing photons produced in the galaxy are assumed to power the
Ly$\alpha$ emission (dashed curve).  However, a higher $f_{\rm esc}$
results in a larger HII region around the galaxy, allowing a larger
fraction of the Ly$\alpha$ line to be transmitted (dotted curve).  The
observed suppression of the line by a factor of upto $\sim 20$ is
consistent with $1\%\lsim f_{\rm esc}\lsim 60\%$.}
\label{fig:fesc}
\end{figure}

\end{document}